\documentclass[useAMS,usenatbib,usegraphicx]{mn2e}
\usepackage{lscape}

\newcommand{\chandra}{{\it Chandra}}
\newcommand{\lum}{\thinspace\hbox{$\hbox{ergs}\thinspace\hbox{s}^{-1}$}}

\title[The Long-term Variability of the X-ray Sources in M82]{The Long-term Variability of the X-ray Sources in M82}
\author[Yi-Kuan Chiang and Albert K.~H.~Kong]{Yi-Kuan Chiang$^{1}$ and Albert K.~H.~Kong$^{2,3}$\\
$^{1}$Institute of Astronomy, National Tsing Hua University, Hsinchu, Taiwan 30013.\\
$^{2}$Institute of Astronomy and Department of Physics, National Tsing Hua University, Hsinchu, Taiwan 30013.\\
$^{3}$Golden Jade Fellow of Kenda Foundation, Taiwan\\}
\begin{document}

%\date{Accepted 1988 December 15. Received 1988 December 14; in original form 1988 October 11}

\pagerange{\pageref{firstpage}--\pageref{lastpage}} \pubyear{2010}

\maketitle

\label{firstpage}

\begin{abstract}
We investigate the long-term variability exhibited by the X-ray point sources in the starburst galaxy M82. By combining 9 \chandra\ observations taken between 1999 and 2007, we detect 58 X-ray point sources within the $D_{25}$ isophote of M82 down to a luminosity of $\sim 10^{37}$\lum. Of these 58 sources, we identify 3 supernova remnant candidates and one supersoft source. Twenty-six sources in M82 exhibit long-term (i.e., days to years) flux variability and 3 show long-term spectral variability. Furthermore, we classify 26 sources as variables and 10 as persistent sources. Among the total 26 variables, 17 varied by a flux ratio of $> 3$ and 6 are transient candidates. By comparing with other nearby galaxies, M82 shows extremely strong long-term X-ray variability that 47\% of the X-ray sources are variables with a flux ratio of $> 3$. The strong X-ray variability of M82 suggests that the population is dominated by X-ray binaries.
\end{abstract}


\section{Introduction}

Based on previous studies of X-ray sources in external galaxies (see Fabbiano and White 2006; Fabbiano 2006), the X-ray populations with luminosity above typical \chandra\ detection limit ($\sim 10^{37}$\lum) of nearby galaxies are dominated by X-ray binaries (XRBs), which consist of an accreting compact object (neutron star or black hole) and a stellar companion. A few young supernova remnants (SNRs) may also be expected. X-ray binaries with low mass stellar companions are called low-mass X-ray binaries (LMXBs), which are usually associated with old stellar populations (bulges, globular clusters and early-type galaxies). On the other hand, high-mass X-ray binaries (HMXBs) with high mass companions are associated with young stellar populations like spiral arms and star-forming regions. Moreover, some off-nuclear ultraluminous X-ray sources (ULXs) with $L_{\rm{X}} > 10^{39}$\lum\
(Eddington luminosity of stellar mass objects) were found in several galaxies, especially star-forming galaxies. The nature of ULXs is still intensely debated. For star-forming galaxies, the source populations are also dominated by XRBs, especially HMXBs (Helfand and Moran 2001). Because of the association between HMXBs and short-lived young stellar populations, the amount of HMXBs may constitute a tracer of recent star formation.

Today's X-ray source population studies in nearby galaxies are mostly based on X-ray luminosity functions (XLFs) and the intensity and spectral properties of individual sources. Furthermore, multiwavelength observations and studies of counterparts also provide important information. X-ray observations of a given galaxy at multiple epochs have provided opportunities to study long-term variabilities of X-ray sources in galaxies. Until now, various kinds of long-term intensity and spectral variability of different types of X-ray sources have been detected in both our own Galaxy and external galaxies. For example, black hole XRBs in our Galaxy are observed in five distinct spectral/temporal states, which are believed to be related to different accretion rates (Esin et al.~1997). Strong flux and spectral variability are shown during state transitions on long timescale. Comparing variability properties of extragalactic X-ray sources with the well studied Galactic X-ray sources can assist in probing the nature of the sources, identifying the sources and understanding the source populations.

In this paper, we report the long-term variability of the X-ray sources in M82 by using 9 archival \chandra\ data sets. M82, a prototypical star-forming galaxy with a distance of 3.6 Mpc (Freedman et al.~1994), exhibits intense star-formation activity and hosts the most luminous ULX, M82 X-1 ($L_{\rm{X}} > 10^{41}$\lum). Although M82 is an important nearby starburst galaxy, not much work has been done on the X-ray source population and the long-term variability of the whole galaxy due to its crowded source distribution, a strong absorption ($N_{\rm{H}} \sim 3\times10^{22}$ cm$^{-2}$) and the intense and non-uniform diffuse emission in the central region of M82. X-ray studies of M82 mostly focused on M82 X-1, the prime candidate of intermediate-mass black hole (IMBH). For the X-ray source population, Zezas et al.~(2001) detected 24 point sources in M82 and 12 of them varied by factors of 0.2 to 7 on timescales of 1 to 6 months. An XLF of the sources was then constructed and can be fitted with a power law with a slope of $\sim -0.4$. Kilgard et al.~(2002) studied the X-ray point sources of three starburst galaxies and four nonstarburst spiral galaxies. They presented an XLF of more than 30 sources detected in M82 with a slope of $\sim -0.5$ and further showed that the XLFs of starburst galaxies tend to be flatter than those of spiral galaxies. Griffiths et al.~(2000) studied the hot plasma and the point sources in M82. They resolved about 20 compact X-ray sources and suggested that many of which could be black hole HMXBs. Now, much more X-ray observations of M82 at multiple epochs have been done, allowing us to study the X-ray population and the long-term variability in detail.

In \S\,2 we describe the \chandra\ observations. We present the data analysis in \S\,3 and the results in \S\,4. A discussion of the results is given in \S\,5.

\section[]{Observations}

Since the X-ray sources in the central region of M82 are crowded, a high spatial resolution instrument like the \chandra\
 X-ray Observatory is needed to resolve the images. M82 has been observed 13 times between the year of 1999 and 2007 with \chandra
; some of them were designed to observe M82 X-1 with special configurations. To study the long-term variability of the whole galaxy, we used 9 \chandra\ data sets with relatively long exposure time and large field-of-view from the \chandra\ Data Archive. The details of these observations are given in Table 1. Among these 9 observations, observation 3, 4, 8 and 9 were performed with the High Resolution Camera (HRC-I or HRC-S). The rest were performed with the Advanced CCD Imaging Spectrometer array (ACIS-I or ACIS-S). For Obs.~ID 1411, there are two separate observations within the same event list. Therefore, we used a time-filter to split them (observation 3 and 4) and analyzed the data separately.

\begin{table}
\centering{\scriptsize
\caption{\chandra\ Observation Log}
\begin{tabular}{cccccc}
\hline
\hline
No. & Obs.~ID & Date & MJD$-$51300 & Exp. & Instrument \\
\hline
		 1 &        361 &  1999/9/20 &     141.47 &     33.7ks &     ACIS-I \\

         2 &       1302 &  1999/9/20 &     141.88 &     15.7ks &     ACIS-I \\

         3 &     1411-1 & 1999/10/28 &     179.18 &     36.3ks &      HRC-I \\

         4 &     1411-2 &  2000/1/20 &     263.64 &     17.8ks &      HRC-I \\

         5 &       2933 &  2002/6/18 &    1143.78 &     18.3ks &     ACIS-S \\

         6 &       5644 &  2005/8/17 &    2299.04 &     75.1ks &     ACIS-S \\

         7 &       6361 &  2005/8/18 &    2300.66 &     19.2ks &     ACIS-S \\

         8 &       8189 &   2007/1/9 &    2809.34 &     61.6ks &      HRC-S \\

         9 &       8505 &  2007/1/12 &    2812.09 &     83.6ks &      HRC-S \\

\hline
\end{tabular}
}
\par
\medskip
\begin{minipage}{0.95\linewidth}
NOTE.--- Obs.~ID 1411 has two observations merged in one event
list. We used a time filter to separate the observations.
\end{minipage}
\end{table}

\section[]{Data Analysis}

The data reduction and analysis were done with CIAO, Version 4.2 (CALDB Version 4.3.0) and HEAsoft, Version 6.3. The ACIS and HRC data sets were processed with the standard procedure by the \chandra\ X-ray Center (CXC). We obtained the fully processed science products (level 2 event files) of these observations. For the ACIS data sets, we only used events between $0.3-7$ keV. In order to remove the differences in the coordinates of these \chandra\ data sets, we registered them using 3 bright X-ray point sources.

\subsection{Source detection}

With WAVDETECT, a wavelet detection algorithm implemented within CIAO, we can search for and determine the position of the point sources. To process the images of the crowded central region, the 0.49 arcsec pixel size of ACIS images is insufficient. In contrast, HRC images have smaller pixels (0.13 arcsec) which oversample the PSF of an on axis point source. Therefore, HRC has the highest spatial resolution of \chandra
. For all the ACIS data sets, we generated the subpixel images with a pixel size of 0.123 arcsec ($=1/4$ ACIS pixel) to reach the resolution of HRC images.

We applied WAVDETECT on all the $0.3-7$ keV $1/4$ ACIS subpixel images and the HRC images at five scales (1, 2, 4, 8 and 16 pixels). We set the detection thresholds to be the inverse of the total pixel number of the images, which is equivalent to stating that the expected number of false sources per field is one. However, there are some spurious detections due to the strong diffuse emission in the central region of M82. After visual inspection, we obtained a candidate source list by combining the WAVDETECT source lists generated for each observation. We used circular source extraction regions of varying radii between $1\arcsec$ to $4\arcsec$ to cover more than 90\% energy of the sources. For each source, a nearby source free region was extracted as background. The background extractions need to be treated especially carefully in the central region since the diffuse emission is not uniform. After intensity and spectral analyses, source candidates with a flux below 1$\sigma$ were first rejected. A source candidate embedded in a relatively irregular part of the diffuse emission will be further considered to be a background contamination if its color is as soft as the diffuse hot gas. In general, sources within the optical $D_{25}$ isophote ($\sim$ $11.6\arcmin$ $\times$ $5.6\arcmin$; $\sim 12$ kpc $\times$ 6 kpc) are thought to be associated with M82. The background contribution was further estimated.

\subsection{Light curves}

To estimate the X-ray flux of each source, we assume an absorbed power-law model with a photon index of 1.7 and a column density of hydrogen $N_{\rm{H}} = 3\times10^{22}$ cm$^{-2}$ (the typical value we obtained from the spectral fittings of the bright sources in M82). We generated the redistribution matrix files (RMFs) and the auxiliary response files (ARFs) of the sources in each observation. Applying the response files and the spectral model described above, we obtained the energy fluxes in $0.3-7$ keV using a tool called modelflux. We further converted these absorbed fluxes into intrinsic fluxes with PIMMS (Portable, Interactive Multi-Mission Simulator). A distance of 3.6 Mpc to M82 (Freedman et al.~1994) is adopted for the estimation of luminosities. If the net counts of a source in an observation is below 1$\sigma$, we obtained the 90\% upper limit of the counts (Gehrels 1986) and the corresponding flux. To deal with the Poisson and binomial statistics inherent to small numbers of counts, all the errors of counts are calculated using Gehrels' error approximation (Gehrels 1986).

\subsection{Hardness ratios}

Since many of the sources are too dim, it is inappropriate to fit the spectra of these faint sources with small degrees of freedom. To obtain a rough picture of spectral properties, we computed the hardness ratios of all the sources in each ACIS data set (observation 1, 2, 5, 6 and 7). The soft color is defined as
\begin{equation}
\textrm{HR1}=\frac{\textrm{M}-\textrm{S}}{\textrm{T}},
\end{equation}
and the hard color as
\begin{equation}
\textrm{HR2}=\frac{\textrm{H}-\textrm{M}}{\textrm{T}},
\end{equation}
where S, M and H are the net counts in the soft ($0.3-1$ keV), medium ($1-2$ keV), and hard ($2-7$ keV) bands, and T are the net counts in the whole energy band($0.3-7$ keV). The errors of hardness ratios were derived from error propagation.

\subsection{Variability}

To estimate long timescale flux variability, we calculated the maximum significance of the difference in fluxes between any two observations,
\begin{equation}
S_\mathrm{flux}=\max_{i,j}\frac{|F_i-F_j|}{\sqrt{\sigma_{F_i}^2+\sigma_{F_j}^2}},
\end{equation}
where $F_i$ is the energy flux of a source in the $i\textrm{th}$ observation and $\sigma_{F_i}$ is the corresponding error. The 90\% upper limit was derived when a source is below 1$\sigma$ detection. $S_\mathrm{flux}$ represents the confidence level of flux variation. In this paper, variations with 3$\sigma$ significance are thought to be detected. However, it is inevitable that the significance parameter of a source is affected by the flux level it reaches, since the fractional error of flux of a faint source is larger than that of a brighter source. In other words, variations of faint sources are hard to be detected even if these sources went through intrinsic variations between observations.

To quantify the intensity of variations, we also computed the ratio of fluxes,
\begin{equation}
R=\frac{F_\mathrm{max}}{F_\mathrm{min}},
\end{equation}
where $F_\mathrm{max}$ and $F_\mathrm{min}$ are the maximum and minimum fluxes of a source among these 9 observations. $R$ could be a number with a standard error if the lowest flux is detected or a $90\%$ lower bound if the lowest flux is an upper limit.

In addition, we checked intra-observation variability for the sources in all the data sets via the Gregory-Loredo algorithm (Gregory and Loredo 1992). Source variability is evaluated from the arrival times of photon events with $m$ bins per observation, where $m$ runs from 2 to $m_{\rm{max}}$. Fractional areas considering the effects of dither motion and bad pixels were corrected. Based on the Gregory-Loredo probability, the odds ratio and the deviation of the light curve from the mean value, an integer variability index could be assigned for the source, which provides a measurement of intra-observation variability in various time scales.

For long-term spectral variability, we computed the significance of the difference in soft color as
\begin{equation}
S_\mathrm{HR1}=\max_{i,j}\frac{|\textrm{HR1}_i-\textrm{HR1}_j|}{\sqrt{\sigma_{\mathrm{HR1}_i}^2+\sigma_{\mathrm{HR1}_j}^2}},
\end{equation}
and the significance of the difference in hard color as
\begin{equation}
S_\mathrm{HR2}=\max_{i,j}\frac {|\textrm{HR2}_i-\textrm{HR2}_j|}{\sqrt{\sigma_{\mathrm{HR2}_i}^2+\sigma_{\mathrm{HR2}_j}^2}}.
\end{equation}
It should be noted that we can obtain spectral information only in 5 observations (observation 1, 2, 5, 6 and 7) which were performed with ACIS-I or ACIS-S. These 5 ACIS observations span over 6 years.

\section[]{Results}

\subsection{Source detection}

After the processes we described in \S\,3, we detected 58 X-ray point sources within the optical $D_{25}$ isophote of M82 in these 9 \chandra\ observations during a span of 7 years. Depending on sources' location and exposure time, the detection limit is $\sim 8\times10^{36}$\lum\ in the outer region and increases to $3\times10^{37}$\lum\ in the central region due to the strong diffuse emission. Among the total 58 X-ray point sources we detected, 25 are in the central $1\arcmin\times1\arcmin$ region and 33 sources are in the outer region with fluxes down to a fainter limit. One of the 25 detections in the central $1\arcmin\times1\arcmin$ region and 6 of the 33 detections in the outer region are expected to be background sources. The background contribution we estimated is based on the results of the \chandra\ deep field north survey (Brandt et al.~2001), which provides an estimate of background sources (AGNs or starburst galaxies) density above a given flux limit.

We also compared the positions of our sources with a list of 37 supernova remnants provided by the deep MERLIN 5GHz radio observation (Fenech et al.~2008), which covers the central 1 kpc region of M82. Two of the X-ray sources (CXOU J095550.7+694044 and CXOU J095552.8+694045) have corresponding radio-selected SNRs located within their positional errors. For the historical type II supernova SN 2004am (Singer et al.~2004; Mattila et al.~2004; Beswick et al.~2004), there is an X-ray source CXOU J095546.7+694038 associated with the optical position.

\begin{table*}
\centering{\scriptsize
\caption{\chandra\ 
source properties for M82}
\begin{tabular}{cccccccccc}
\hline
\hline
No. & Source Name & R.~A. & Decl. & Max. $L_{\rm{X}}$$^{a}$ & $S_\mathrm{flux}$$^{b}$ & $R$$^{c}$ & $S_\mathrm{HR1}$$^{d}$ & $S_\mathrm{HR2}$$^{e}$ & Notes$^{f}$ \\
 &  (CXOU J) & \multicolumn{2}{c}{ (J2000)} &  ($10^{38}$\lum) & & & & & \\ 
\hline

1 & 095510.7+693955 &  148.79455 &  69.665369 & 2.68 & 2.78 & 3.64$\pm$1.41 & 0.58 & 1.2 & P \\ 
2 & 095527.2+693923 &  148.86338 &  69.656503 & 11.71 & 5.59 & 2.23$\pm$0.3 & 1.5 & 1.67 & V \\ 
3 & 095527.2+694050 &  148.86345 &  69.680558 & 0.58 & $>$2.58 & $>$8.48 & ... & ... &  \\ 
4 & 095529.1+694027 &  148.87116 &  69.674172 & 0.88 & $>$2.68 & $>$6.45 & 1.24 & 1.41 &  \\ 
5 & 095529.2+694212 &   148.87180 &  69.703353 & 1.14 & $>$1.15 & $>$3.09 & 0.78 & 0.35 &  \\ 
6 & 095530.2+694239 &   148.87570 &  69.710836 & 0.81 & $>$0.85 & $>$2.37 & 0.07 & 0.41 &  \\ 
7 & 095530.9+694234 &  148.87886 &  69.709403 & 1.15 & $>$1.68 & $>$5.22 & 0.41 & 0.4 &  \\ 
8 & 095531.9+693958 &   148.88280 &  69.665981 & 0.87 & $>$1.45 & $>$3.98 & 0.43 & 0.52 &  \\ 
9 & 095532.7+694001 &  148.88633 &  69.667075 & 2.98 & 4.35 & 21.5$\pm$14.93 & 0.82 & 0.29 & V3 \\ 
10 & 095534.2+693943 &  148.89275 &  69.662022 & 0.85 & $>$1.59 & $>$4.90 & 0.33 & 0.53 &  \\ 
11 & 095534.5+693824 &  148.89392 &  69.639917 & 0.95 & $>$1.51 & $>$2.77 & 0.93 & 0.33 &  \\ 
12 & 095537.9+694057 &  148.90792 &  69.682597 & 0.90 & $>$1.64 & $>$7.85 & 0.56 & 1.16 &  \\ 
13 & 095538.0+694030 &  148.90853 &  69.674875 & 6.02 & 4.58 & 2.16$\pm$0.32 & 1.19 & 1.64 & V \\ 
14 & 095541.0+693928 &  148.92068 &  69.657761 & 2.76 & $>$4.43 & $>$25.41 & ... & ... & V3 \\ 
15 & 095542.8+694033 &   148.92850 &  69.675839 & 0.81 & $>$1.18 & $>$2.81 & 0.4 & 0.82 &  \\ 
16 & 095544.1+693958 &  148.93383 &  69.666206 & 0.61 & $>$1.18 & $>$2.74 & 0.47 & 0.23 &  \\ 
17 & 095546.2+694026 &  148.94247 &  69.673997 & 5.47 & 2.79 & 1.75$\pm$0.31 & 2.32 & 1.14 & P \\ 
18 & 095546.6+694041 &  148.94403 &  69.677994 & 98.16 & 37.57 & 4.69$\pm$0.36 & 3.56 & 5.07 & V3, SV, IOV(6,7) \\ 
19 & 095546.7+694038 &  148.94453 &  69.677175 & 5.72 & 5.89 & 2.34$\pm$1.03 & 0.98 & 0.86 & SN 2004am, V \\ 
20 & 095547.4+694036 &  148.94768 &  69.676733 & 46.06 & $>$21.18 & $>$884.19 & ... & ... & T, IOV(3) \\ 
21 & 095547.5+694100 &  148.94777 &  69.683258 & 4.31 & 2.68 & 1.45$\pm$0.27 & 1.07 & 1.72 & P \\ 
22 & 095548.8+694044 &  148.95328 &  69.678783 & 0.84 & $>$0.61 & $>$1.64 & 0.79 & 0.02 & IOV(7) \\ 
23 & 095549.4+694044 &  148.95594 &  69.678783 & 2.88 & 1.41 & 1.99$\pm$1.23 & 1.68 & 1.12 & P, IOV(8) \\ 
24 & 095550.0+693934 &  148.95822 &  69.659514 & 4.34 & 6.41 & 5.31$\pm$1.39 & 0.97 & 1.66 & V3 \\ 
25 & 095550.1+694046 &  148.95889 &  69.679569 & 1581.69 & 181.28 & 28.25$\pm$0.82 & 8.75$^{g}$ 3.47$^{h}$ & 8.37$^{i}$ 4.76$^{j}$ & M82 X-1, V3, SV, IOV(6) \\ 
26 & 095550.3+694022 &  148.95978 &  69.672908 & 8.36 & 5.77 & 2.84$\pm$0.51 & 0.52 & 1.58 & V \\ 
27 & 095550.4+694036 &  148.95987 &  69.676767 & 11.34 & $>$15.84 & $>$78.22 & 0.31 & 0.87 & T \\ 
28 & 095550.7+694044 &  148.96105 &  69.678783 & 20.62 & 5.39 & 1.75$\pm$0.25 & 2.18 & 3.12 & RSN, V, SV \\ 
29 & 095551.0+694045 &  148.96243 &  69.679194 & 108.98 & 98.6 & 65.7$\pm$25.61 & 1.15 & 2.21 & T \\ 
30 & 095551.3+694044 &  148.96361 &  69.678853 & 35.63 & 12.28 & 2.85$\pm$0.31 & 3.17 & 1.64 & V \\ 
31 & 095551.5+694036 &   148.96450 &  69.676667 & 22.33 & 5.79 & 1.93$\pm$0.31 & 0.93 & 1.72 & V \\ 
32 & 095551.9+694050 &  148.96607 &  69.680458 & 1.36 & $>$1.97 & $>$6.15 & 1.91 & 1.41 &  \\ 
33 & 095552.0+694043 &  148.96647 &  69.678614 & 4.90 & $>$4.93 & $>$3.99 & 0.8 & 1.17 & V3 \\ 
34 & 095552.3+694054 &  148.96774 &  69.681586 & 24.27 & $>$24.59 & $>$87.42 & 1.12 & 2.68 & T \\ 
35 & 095552.6+694047 &  148.96932 &  69.679808 & 3.80 & 2.85 & 5.62$\pm$4.75 & 1.36 & 2.81 & P \\ 
36 & 095552.8+694045 &  148.96981 &  69.679297 & 5.43 & 2.81 & 4.25$\pm$4.11 & 1.73 & 1.27 & RSN, P \\ 
37 & 095553.2+694049 &  148.97168 &  69.680219 & 5.50 & 4.11 & 4.19$\pm$1.89 & 1.58 & 1.92 & V3 \\ 
38 & 095553.4+694102 &  148.97256 &  69.683875 & 2.20 & 1.81 & 2.18$\pm$1.16 & 0.97 & 1.02 & P \\ 
39 & 095553.4+693956 &  148.97257 &  69.665461 & 2.54 & 2.12 & 3.35$\pm$3.16 & 1.25 & 0.8 & P, IOV(2) \\ 
40 & 095553.8+694050 &  148.97414 &  69.680664 & 9.99 & $>$8 & $>$7.77 & 1.73 & 1.44 & V3 \\ 
41 & 095554.2+694040 &  148.97571 &  69.677692 & 10.45 & 11.24 & 11.91$\pm$5.26 & 2.24 & 3.81$^{k}$ & V3 \\ 
42 & 095554.7+694101 &  148.97787 &  69.683569 & 10.65 & 6.29 & 2.19$\pm$0.27 & 1.59 & 1.41 & V \\ 
43 & 095555.1+694028 &  148.97945 &  69.674481 & 1.23 & $>$1.14 & $>$3.07 & 0.94 & 0.24 &  \\ 
44 & 095600.3+694046 &  149.00139 &    69.679400 & 1.26 & $>$2.8 & $>$12.46 & 0.32 & 0.07 & IOV(9) \\ 
45 & 095601.3+694111 &  149.00553 &  69.686469 & 18.30 & $>$17.74 & $>$112.17 & ... & ... & T \\ 
46 & 095601.7+694309 &  149.00705 &  69.719167 & 1.11 & $>$1.63 & $>$2.94 & 0.5 & 0.75 &  \\ 
47 & 095606.0+694031 &    149.02500 &  69.675328 & 0.97 & $>$1.91 & $>$11.59 & 0.66 & 0.16 &  \\ 
48 & 095608.1+694140 &   149.03360 &  69.694322 & 5.94 & $>$8.84 & $>$126.36 & ... & ... & T \\ 
49 & 095614.8+694249 &  149.06164 &  69.713617 & 1.10 & $>$3.72 & $>$9.16 & 0.65 & 0.48 & V3 \\ 
50 & 095615.1+693952 &  149.06283 &  69.664347 & 1.25 & $>$1.67 & $>$3.42 & 0.26 & 0.67 &  \\ 
51 & 095615.2+694142 &  149.06334 &  69.695097 & 1.09 & $>$4.01 & $>$13.71 & ... & ... & SSS, V3 \\ 
52 & 095619.7+694146 &  149.08225 &  69.696147 & 6.89 & 3.71 & 1.91$\pm$0.29 & 0.76 & 1.06 & V \\ 
53 & 095621.2+694225 &  149.08821 &  69.706939 & 1.35 & 1.69 & 3.37$\pm$1.74 & 0.43 & 0.29 & P \\ 
54 & 095621.4+693901 &   149.08920 &  69.650153 & 0.65 & $>$0.32 & $>$1.54 & 0.21 & 0.29 &  \\ 
55 & 095622.7+694135 &  149.09463 &  69.692961 & 1.13 & $>$1.85 & $>$14.26 & 0.41 & 0.01 &  \\ 
56 & 095631.1+694220 &  149.12977 &  69.705611 & 2.88 & $>$2.47 & $>$2.87 & 1.01 & 0.91 &  \\ 
57 & 095633.8+694352 &   149.14080 &  69.731192 & 1.60 & $>$1.65 & $>$2.92 & 0.71 & 0.56 &  \\ 
58 & 095637.6+694137 &  149.15686 &  69.693731 & 1.25 & 1.34 & 1.62$\pm$1.05 & 0.87 & 1.19 & P \\ 
     
\hline
\end{tabular}
}
\par
\medskip
\begin{minipage}{0.95\linewidth}
$^a$ The highest luminosity in 0.3 -- 7 keV of a source among the 9 observations we used, assuming
an absorbed power-law model with a photon index of 1.7 , $N_{\rm{H}} = 3\times10^{22}$ cm$^{-2}$ and d=3.6 Mpc.\\
$^b$ $S_\mathrm{flux}=\max_{i,j}( |F_i-F_j|/\sqrt{\sigma_{F_i}^2+\sigma_{F_j}^2} )$\\
$^c$ $R=F_\mathrm{max}/F_\mathrm{min}$ with a standard error if the lowest flux is detected or a $90\%$ lower bound if the lowest flux is an upper limit.\\
$^d$ $S_\mathrm{HR1}=\max_{i,j}( |\textrm{HR1}_i-\textrm{HR1}_j|/\sqrt{\sigma_{\mathrm{HR1}_i}^2+\sigma_{\mathrm{HR1}_j}^2} )$ of the 5 \chandra\ 
ACIS observations.\\
$^e$ $S_\mathrm{HR2}=\max_{i,j}( |\textrm{HR2}_i-\textrm{HR2}_j|/\sqrt{\sigma_{\mathrm{HR2}_i}^2+\sigma_{\mathrm{HR2}_j}^2} )$ of the 5 \chandra\ 
ACIS observations.\\
$^f$ P: persistent source, V: variable, V3: variable with flux ratio $>3$, T: transient candidate, SV: spectral variable, SSS: supersoft source candidate, RSN: source coincides with a radio-selected supernova remnant in the list of Fenech et al.(2008), IOV(Obs. No.): intra-observation variability detected in the specified observation.\\
$^g$, $^i$ The $S_\mathrm{HR1}$ \& $S_\mathrm{HR2}$ with a different extraction of colors in observation 5 using a pile-up spectral fitting. This source suffers from mild pile-up in observation 5 with a pile-up fraction of $\sim 26\%$.\\
$^h$, $^j$ The $S_\mathrm{HR1}$ \& $S_\mathrm{HR2}$ excluding observation 5.\\
$^k$ Considering the strong contamination from the diffuse emission, this significant level is overestimated.\\
\end{minipage}
\vspace{3mm}
\end{table*}

\subsection{Variability}
As we described in \S\,3.2, the $0.3-7$ keV unabsorbed fluxes were obtained by assuming an absorbed power-law model with a photon index of 1.7 and a column density $N_{\rm{H}} = 3\times10^{22}$ cm$^{-2}$. We further evaluated the variation of the derived unabsorbed fluxes with different spectral parameters. Varying photon index from 1.7 to 1.2 and 2.0 for the sources in observation 1 results in flux changes within 20\%. Combined with the fact that most of the sources in M82 do not undergo spectral variation significantly (details will be presented in this section), the long-term variability of the sources is insensitive to the assumed spectral parameters.

Based on the significance parameter of long-term flux variability $S_\mathrm{flux}$ defined in \S\,3.4, we classify the X-ray point sources into persistent sources if $S_\mathrm{flux} < 3$ or variables if $S_\mathrm{flux} > 3$. For some faint sources with fluxes close to the detection limit, we can only obtain the lower limit of their $S_\mathrm{flux}$. Therefore, it would be impossible for us to classify a source if its lower limit of $S_\mathrm{flux} < 3$. To indicate the amount of variability, we also mark the variables whose flux ratio $R > 3$. Furthermore, if a variable showed a strong variability with $R > 50$, we consider it to be a transient candidate. To describe the long-term variability of the whole galaxy, we use the fraction of variables with $R > 3$ (excluding unclassified sources) to be an indicator.

Among the total 58 x-ray point sources in M82, there are 10 persistent sources, 26 variables and 22 unclassified sources. Seventeen of the 26 variables showed a variability with $R > 3$. Six of the 26 variables are transient candidates. Excluding the unclassified sources, 47\% of the detected X-ray sources in M82 are variables with $R > 3$ on timescale of months-years. The long-term variability properties of the sources in M82 are given in Table 2.

For the central $1\arcmin\times1\arcmin$ region of M82, the population of long-term flux variability is similar to that of the whole galaxy.

Intra-observation variability plays a limited role in the detected flux variances between observations. Sources with Gregory-Loredo variability index less than 3 are considered to be persistent within observation. Of all the sources in M82, only 6 and 1 showed short-term variability in 1 and 2 observations respectively. We note these sources with corresponding observations in Table 2.

Furthermore, we estimated the spectral variability using the significance of the difference in hardness ratios. We consider a source to be a spectral variable if $S_\mathrm{HR1} > 3$ or $S_\mathrm{HR2} > 3$. Originally, we found 5 sources satisfied these criteria but 2 spurious cases have been dismissed. Spurious variation could be detected in a transient like source embedded in a region of strong and irregular diffuse emission. Although the high error of flux when a source was faint should suppress the confidence of its spectral variation, contamination from diffuse emission could be relatively serious in its faint states. Then, a bright outburst could make the composite error small compared to the color difference and a spurious detection may be selected. If a source suffered from strong contamination of the diffuse emission, we expect that the error should be greater than the standard error, and the detection slightly over our criteria should be dismissed. After all, 3 sources exhibited long-term spectral variability with $S_\mathrm{HR1} > 3$ or $S_\mathrm{HR2} > 3$. All of these 3 spectral variables also show variability in their X-ray flux.

Due to the limiting statistics, it is inevitable that the transient candidates and the spectral variables we found tend to be luminous. Sources with relatively low flux and sources within the region of strong diffuse emission hardly reach our criteria of transient candidates and spectral variables due to this selection effect.

\begin{figure*}
\centering
\includegraphics[width=60mm,angle=270]{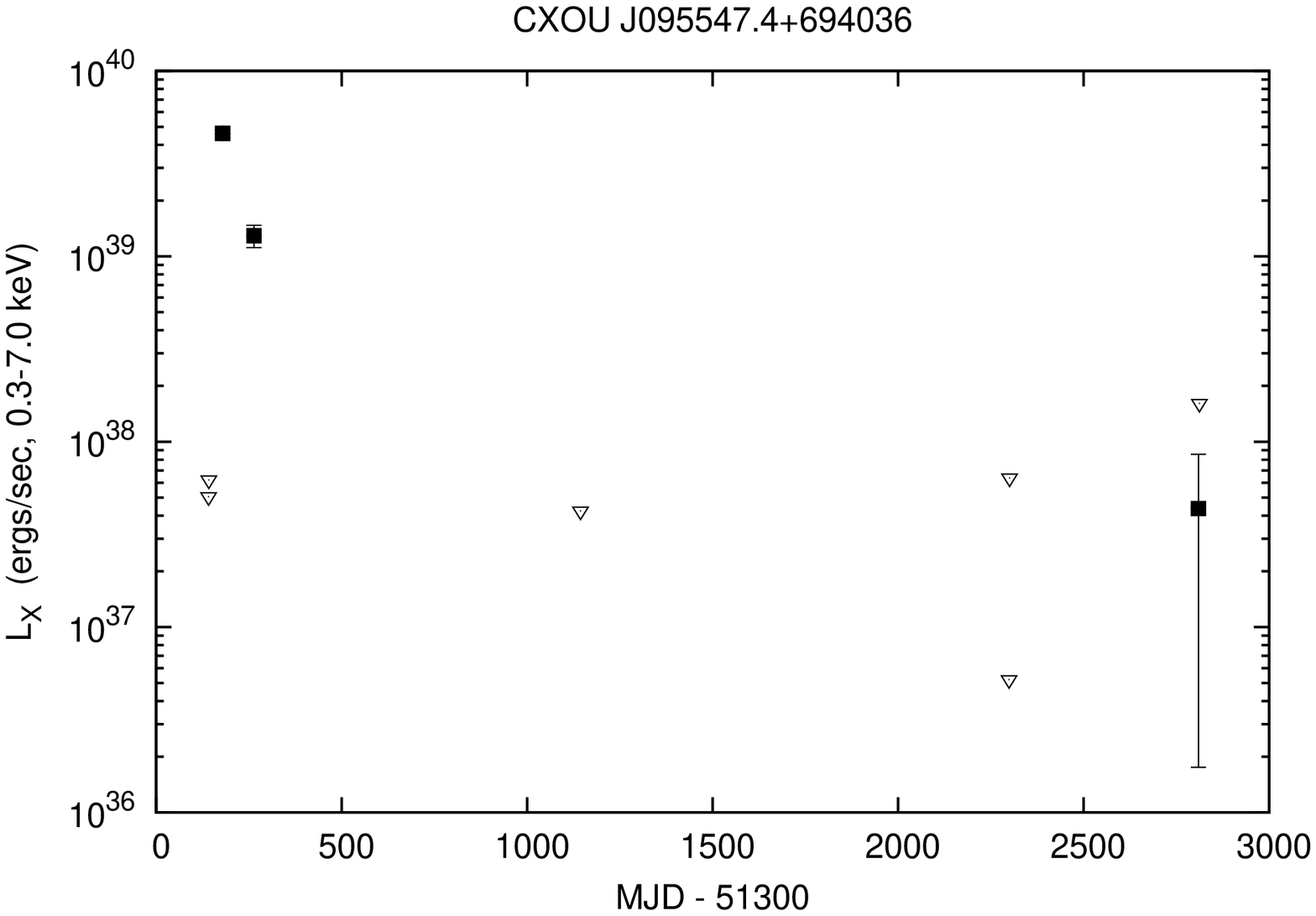}
\includegraphics[width=60mm,angle=270]{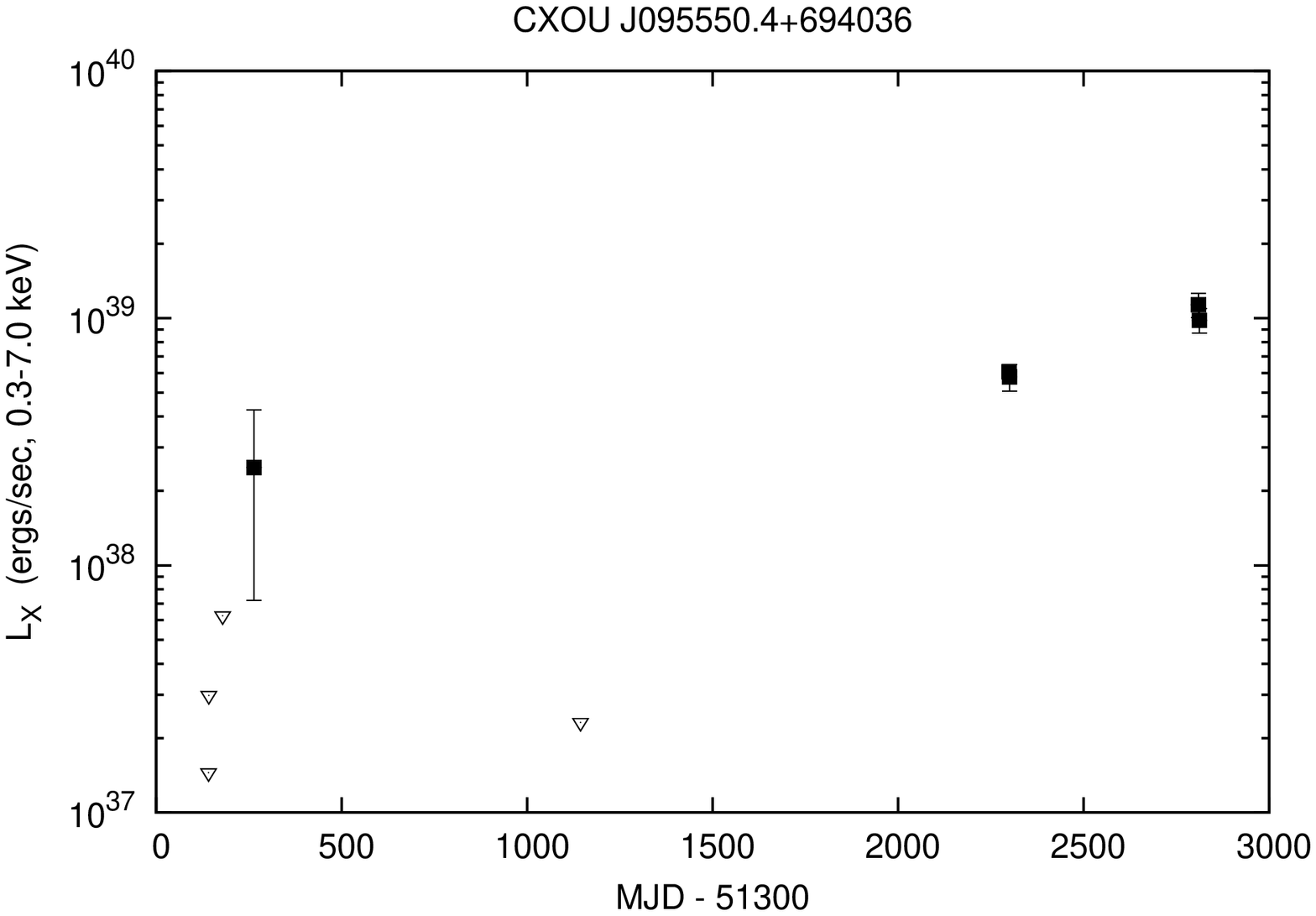}
\includegraphics[width=60mm,angle=270]{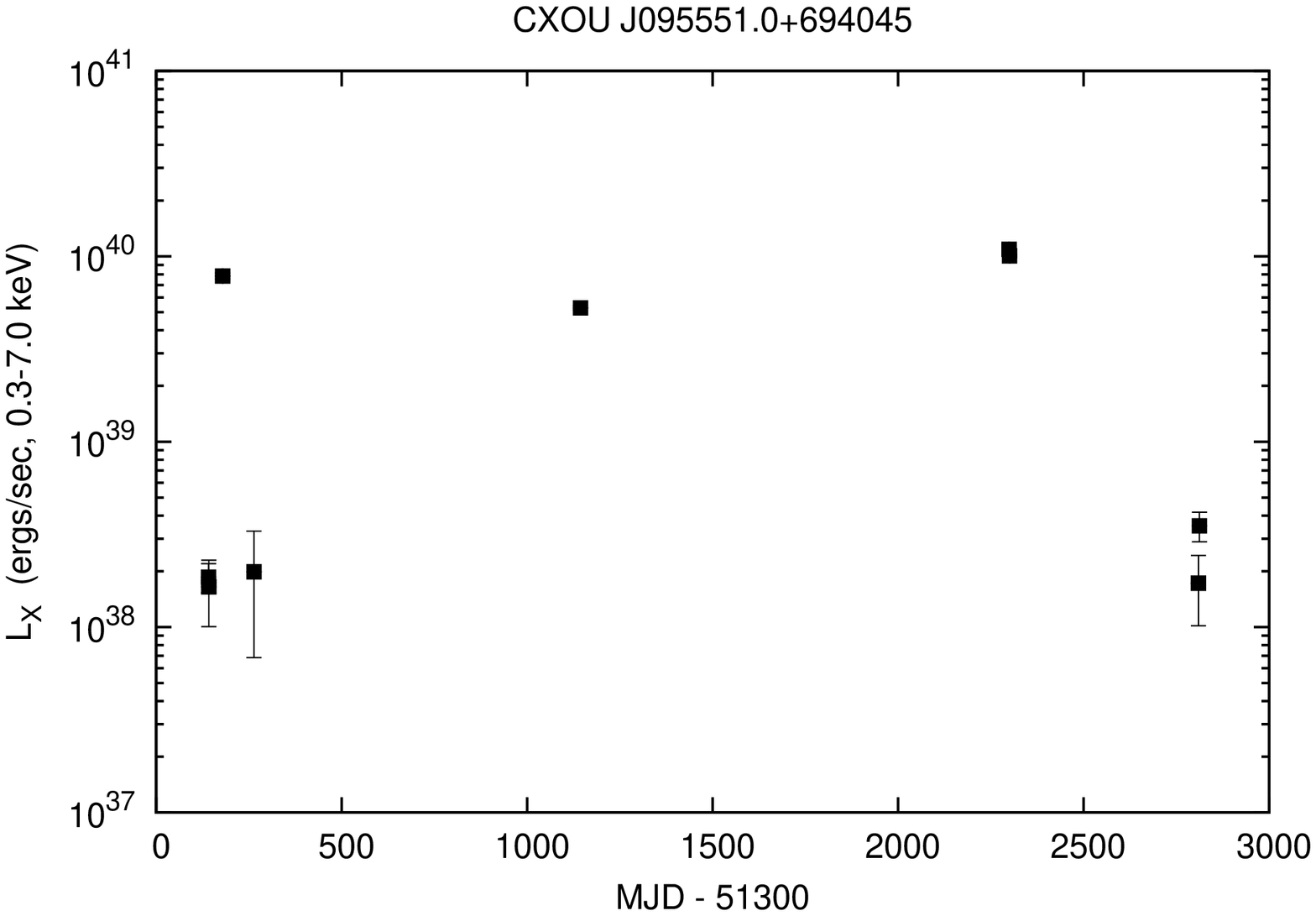}
\includegraphics[width=60mm,angle=270]{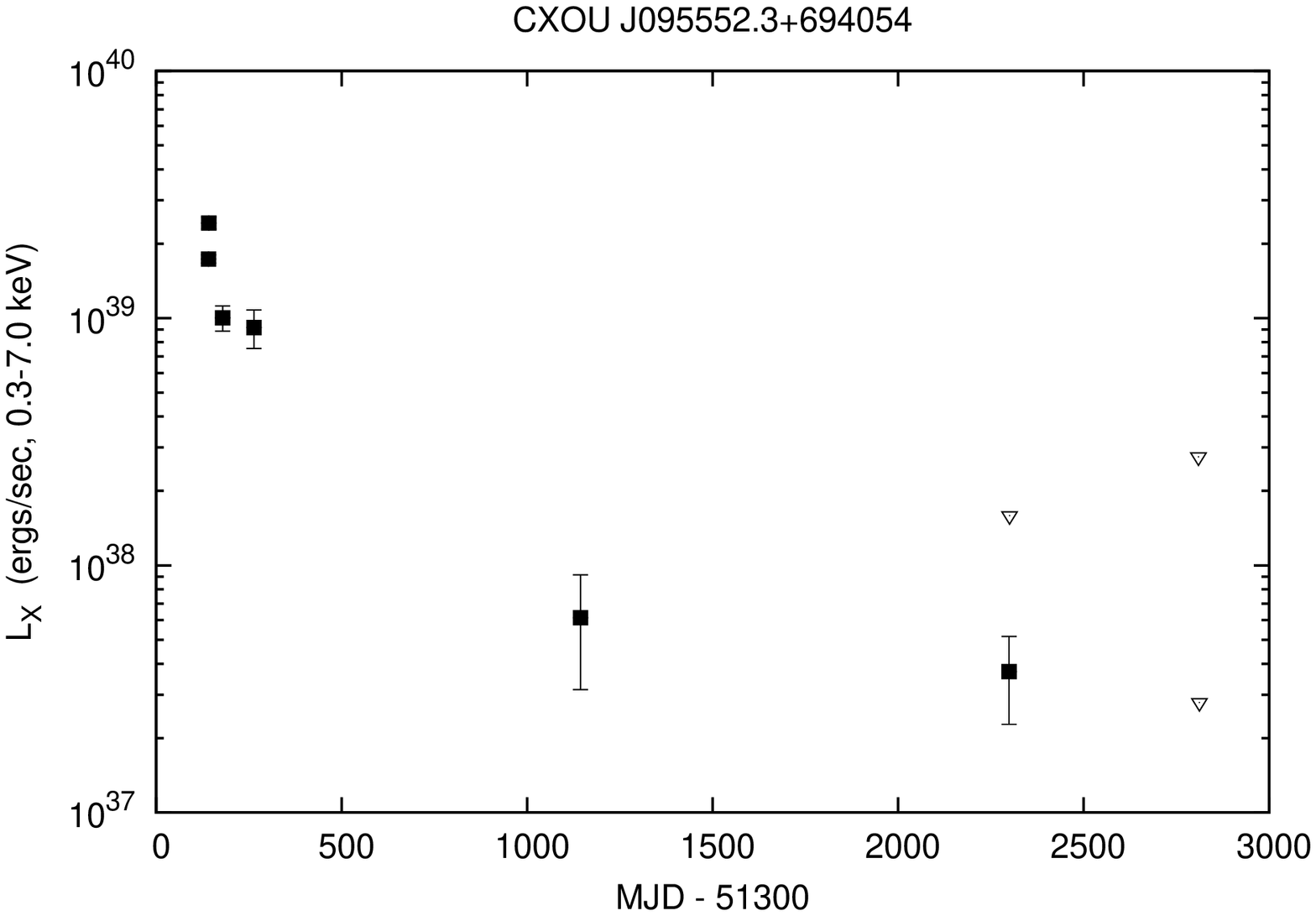}
\includegraphics[width=60mm,angle=270]{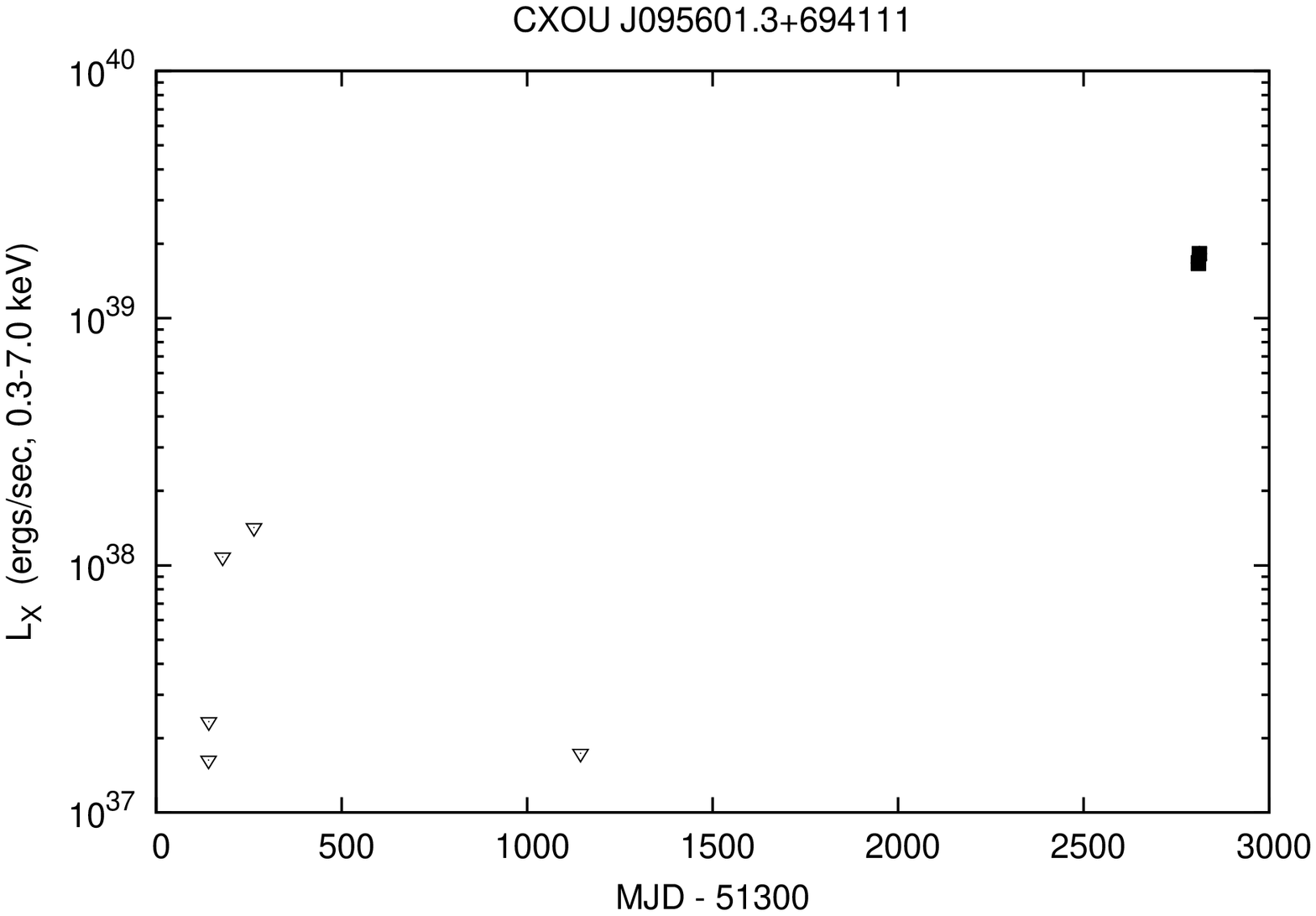}
\includegraphics[width=60mm,angle=270]{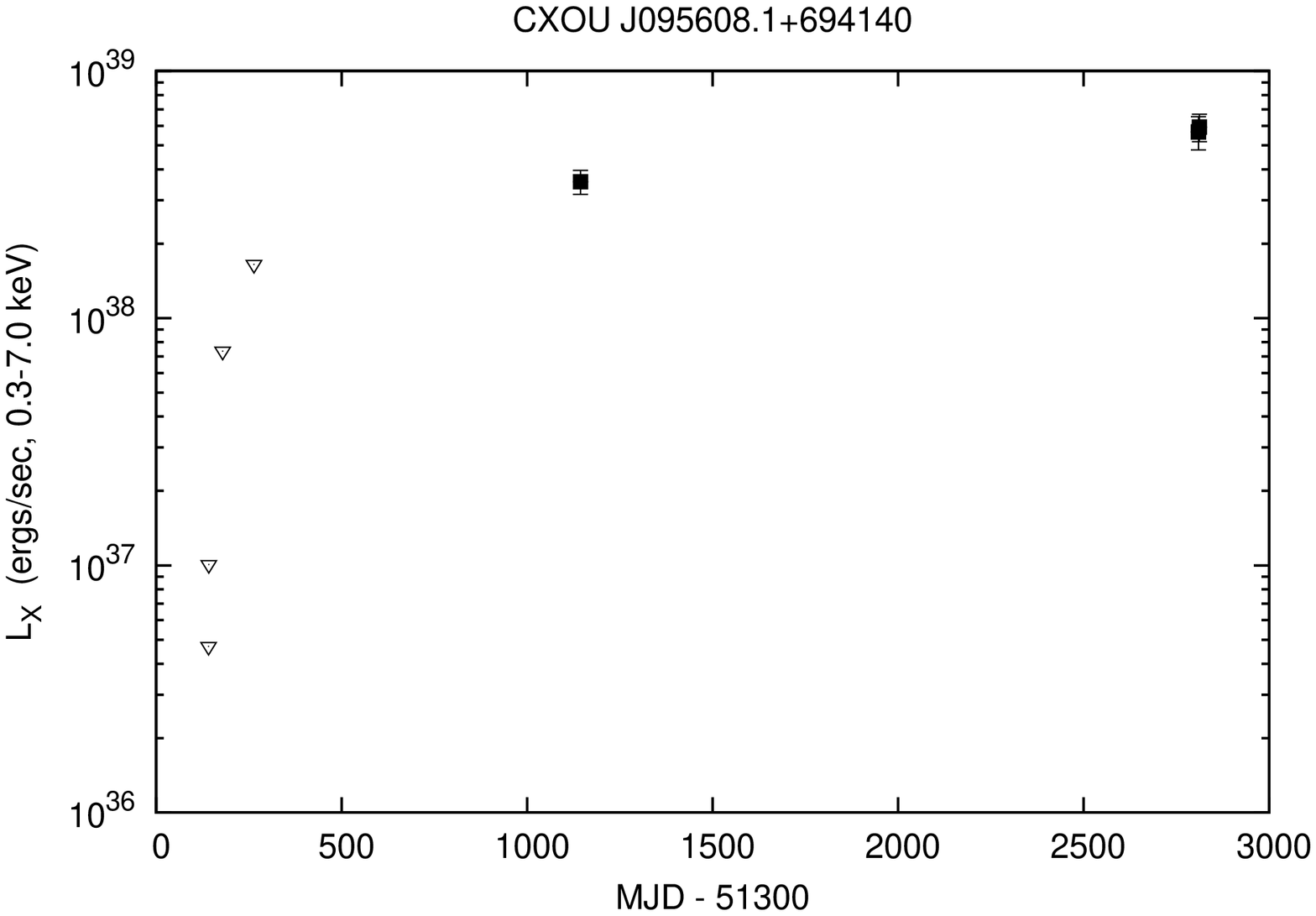}
\caption{Long-term X-ray lightcurve of the transient candidates. The luminosities are determined by assuming
an absorbed power-law model with a photon index of 1.7 , $N_{\rm{H}} = 3\times10^{22}$ cm$^{-2}$ and d=3.6 Mpc. For non-detections, downward triangles represent their 90\% upper limits.}
\end{figure*}

\subsection{Individual sources}

For the sources which are classified as variables, transient candidates or spectral variables, various kinds of variability properties are shown. Some of them may be related to spectral/temporal state transitions of black hole XRBs. However, it is usually hard to confirm because of the lack of low state spectra with our detection limit. X-ray power spectra and studies of radio counterparts are also very useful to probe the nature of the sources.

Some of the transient candidates exhibited outburst once during these 9 \chandra\ observations. One of the transients showed a lightcurve which is consistent with a fast rise followed by an exponential decay (FRED). Two transients showed recurrent outbursts on long timescale. In addition to the 6 transient candidates, some of our sources also showed transient-like behavior but did not meet our criteria for being classified as a transient candidate. We also found a supersoft source (SSS), which only appeared in one observation with all its counts in the soft band and was below detection limit in the other observations. For the 3 spectral variables, we constructed the color-luminosity diagrams. The detailed properties of the individual sources will be presented in this section.

\subsubsection{Transients}

CXOU J095547.4+694036, an X-ray transient candidate which showed a long-term lightcurve consistent with a fast rise followed by an exponential decay (FRED), a typical profile of X-ray novae. This source reached its peak luminosity ($4.6\times10^{39}$\lum) in observation 3 and followed by a lower luminosity of $1.3\times10^{39}$\lum\ three months later. The detailed profile of its flux decay is unclear due to insufficient data points. We do not have spectral information because the 2 observations which caught the outburst were performed with HRC-I.

\begin{figure*}
\centering
\includegraphics[width=6cm,angle=270]{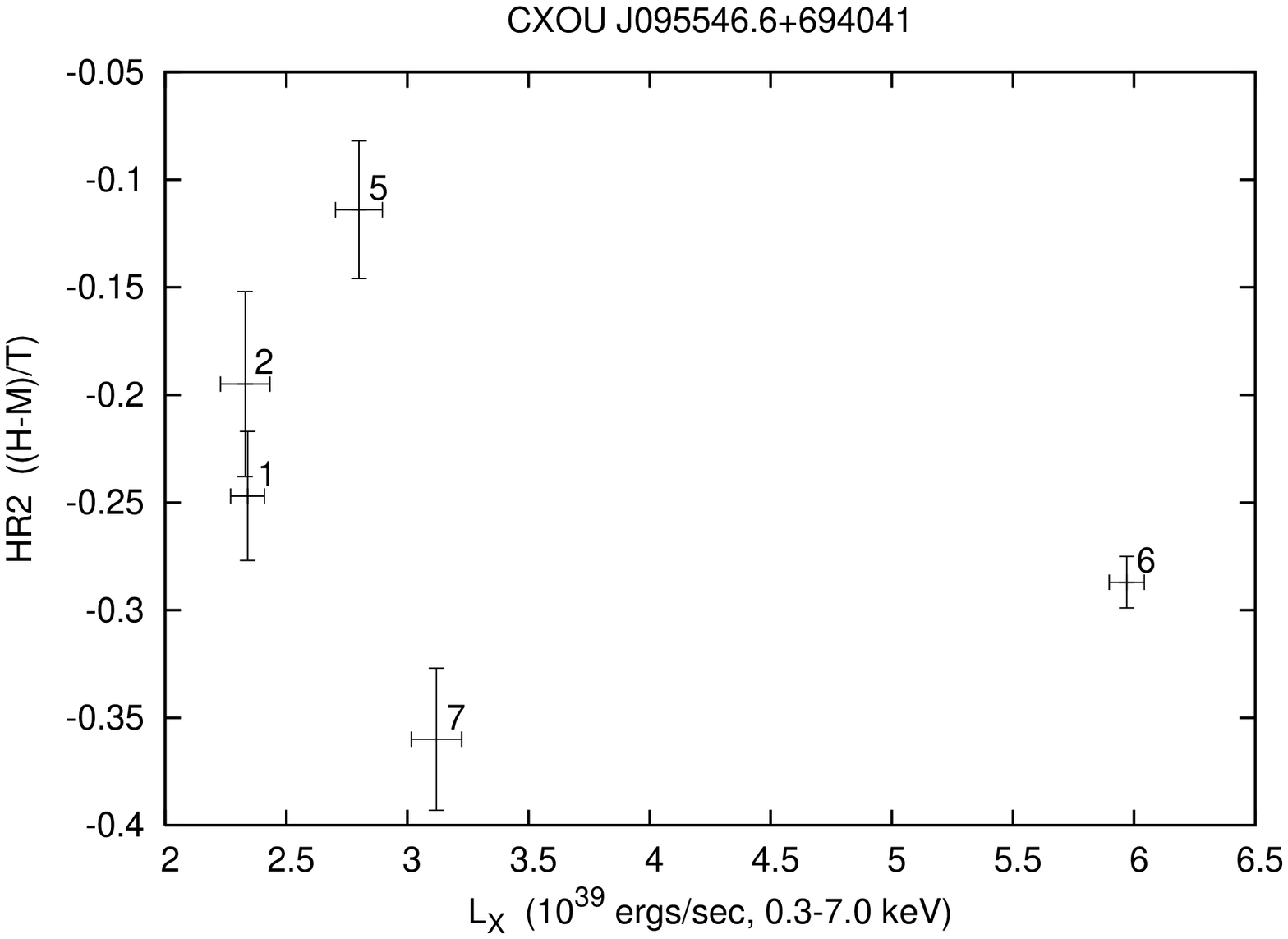}
\includegraphics[width=6cm,angle=270]{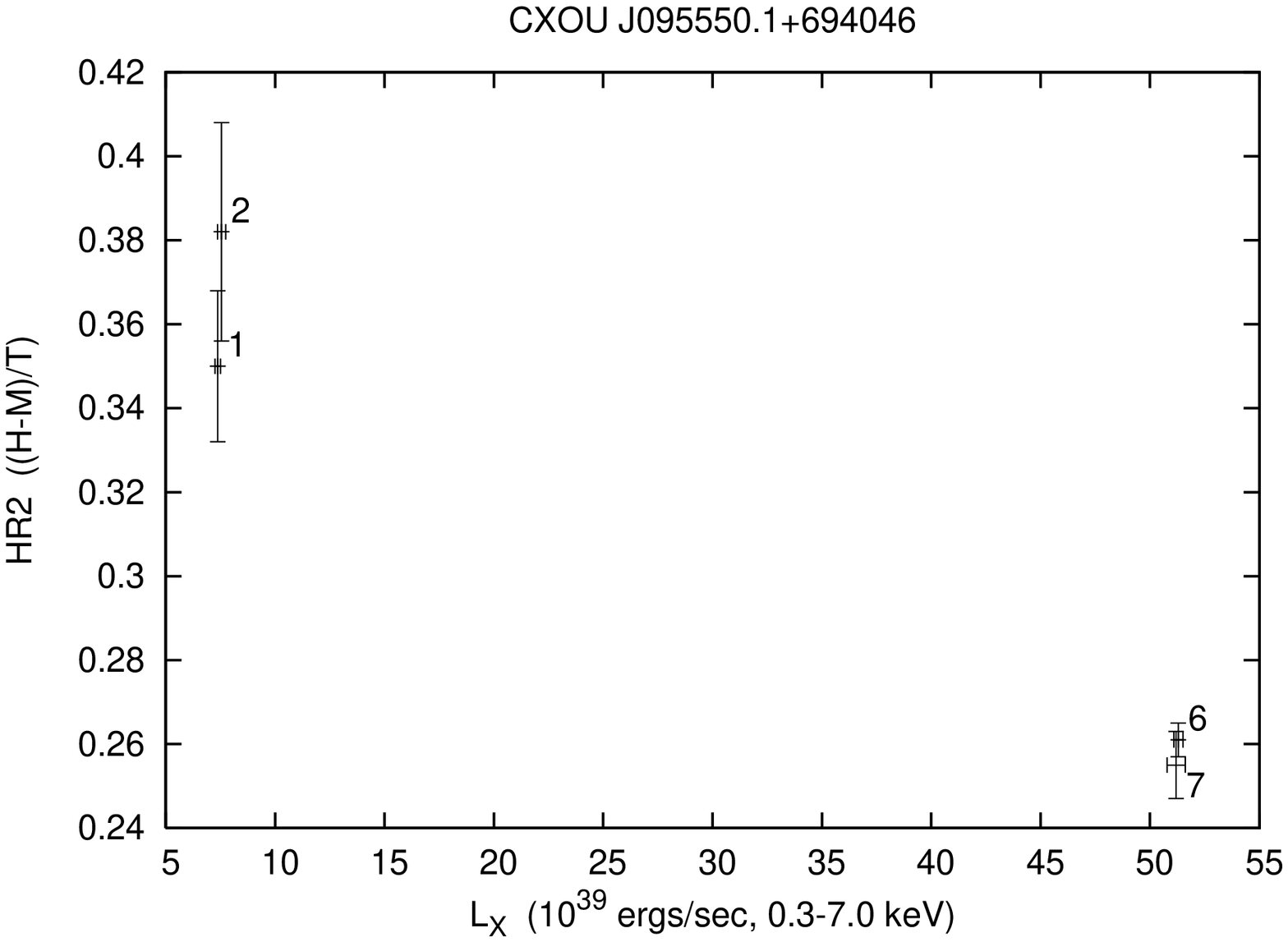}
\includegraphics[width=6cm,angle=270]{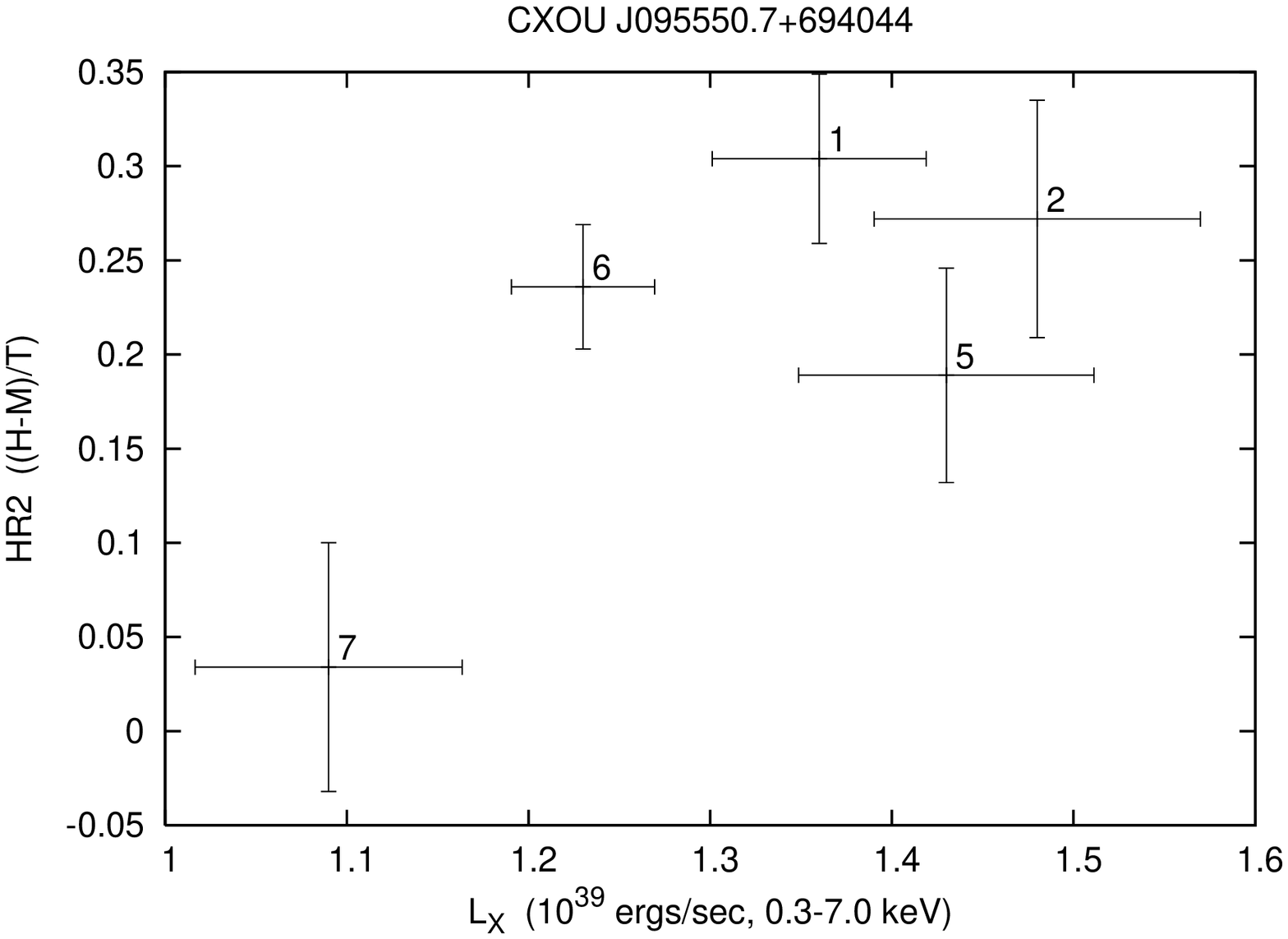}
\caption{Hard color as a function of luminosity for sources that show significant variation in their hardness ratios and are classified as spectral variables. The numbers next to the data points indicate the Obs. No.}
\end{figure*}

CXOU J095552.3+694054, an X-ray transient candidate which was luminous in the first two observations with $L_{\rm{X}} > 10^{39}$\lum. And then it gradually decayed to the detection limit on time scale of years. It is not clear whether it has a fast rise profile or not. The spectra of CXOU J095552.3+694054 in observation 1 can be well fitted with an absorbed power-law model with a photon index of 1.8.

CXOU J095601.3+694111, another transient candidate, originally showed luminosities below the detection limit. However, it appeared in the last two HRC observations with $L_{\rm{X}} > 10^{39}$\lum\ and no spectral information is available. The profile of the outburst is not clear. This source is not covered by the field-of-views of observation 6 and 7.

CXOU J095608.1+694140, an X-ray transient candidate suddenly appeared in observation 5 with $L_{\rm{X}} > 10^{38}$\lum\ and seems to be stable in flux hereafter. The X-ray colors of CXOU J095608.1+694140 also did not vary after observation 5.

The other two transient candidates we found in M82, CXOU J095550.4+694036 and CXOU J095551.0+694045 showed recurrent outbursts between 1999 and 2007. One of the most luminous X-ray sources near the center of M82, CXOU J095551.0+694045, has been studied using roughly the same observations in the work of Kong et al.~(2007). The X-ray luminosity of CXOU J095551.0+694045 reaches $10^{40}$\lum\ in outbursts. When it turned into quiescence in observation 1, 2, 4, 8 and 9, basically it cannot be detected by WAVDETECT. The detected luminosities ($\sim 10^{38}$\lum) in observation 1, 2, 4, 8 and 9 that we present in the lightcurve (Figure 1) may due to contamination of the diffuse emission and nearby luminous point sources. Indeed, the detection limits for this position are just about $10^{38}$\lum. Our lightcurve of this ultraluminous X-ray transient is consistent with which shown in Kong et al.~(2007). For the other two recurrent X-ray transients, one of them (CXOU J095550.4+694036) has spectra available in ACIS observations. The spectra can be fitted with an absorbed power-law with a photon index of $\sim$ 3.7 and a reduced $\chi^2$ of 0.9, indicating that the source is in high/soft state.

In addition to these 6 transient candidates, some dim sources may reach our criteria of transient candidates if there was no diffuse emission. One case is CXOU J095553.8+694050, a recurrent source located in the strong diffuse emission region. Another transient-like source, CXOU J095615.2+694142, is the only SSS we found in M82. It only appeared in observation 5 with its total counts of 30 all in the soft band. The colors and the transient nature of this source are very similar to that of typical SSSs in nearby galaxies (Di Stefano and Kong 2003). We also estimated the luminosity of the SSS by assuming a blackbody model with temperatures of 50 eV and 100 eV. The estimated luminosity strongly depends on the assumed column density of hydrogen. The accurate column density should be located between the value of the Milky Way absorption in this direction ($N_{\rm{H}} = 4\times10^{20}$ cm$^{-2}$, Dickey and Lockman, 1990) and the typical absorption ($N_{\rm{H}} = 3\times10^{22}$ cm$^{-2}$) obtained from the spectral fitting of the bright sources in the central region of M82. With the Galactic absorption ($N_{\rm{H}} = 4\times10^{20}$ cm$^{-2}$) and a blackbody model with a temperature of 50 eV, the absorbed and unabsorbed 0.3-7 keV luminosities are $1.90\times10^{37}$ and $3.90\times10^{37}$ \lum, respectively. For a blackbody model with a temperature of 100 eV, the absorbed and unabsorbed 0.3-7 keV luminosities are $1.27\times10^{37}$ and $1.99\times10^{37}$ \lum, respectively. If we assume a higher typical absorption of the central bright sources in M82 ($N_{\rm{H}} = 3\times10^{22}$ cm$^{-2}$), the expected unabsorbed 0.3-7 keV luminosity of this SSS could reach $7.68\times10^{43}$ \lum\ using a blackbody model with a temperature of 50 eV and $5.21\times10^{40}$ \lum\ for a black body temperature of 100 eV. If the absorption is indeed much higher than the Galactic one, CXOU J095615.2+694142 should be classified as an ultraluminous SSS like the one in M101 (Kong et al.~2004; Kong and Di Stefano 2005).

\begin{figure}
\centering
\includegraphics[width=6.2cm,angle=270]{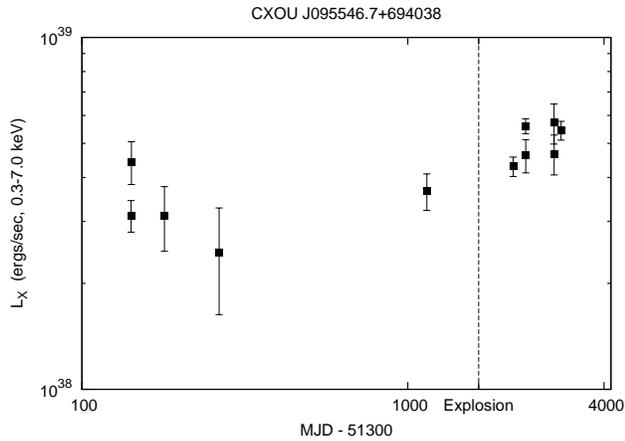}
\caption{Long-term lightcurve of CXOU J095546.7+694038, an X-ray source at the position of a historical supernova SN 2004am. In the observations after the supernova explosion (MJD$-51300\sim 1650$), the X-ray luminosity of it increased by a factor of $\sim$ 50\%. The luminosities are determined by assuming
an absorbed power-law model with a photon index of 1.7 , $N_{\rm{H}} = 3\times10^{22}$ cm$^{-2}$ and d=3.6 Mpc.
}
\end{figure}

\subsubsection{Spectral variables}

M82 X-1, CXOU J095550.1+694046, showed strong X-ray spectral variability in the data set of observation 5 due to pile-up effect. We used the pile-up model to fit the spectra and extracted the hardness ratios for M82 X-1 in observation 5. However, medium pile-up with a pile-up fraction of $\sim 26\%$ from which it suffered could still impair the quality of spectral analysis. Excluding observation 5, the significance parameter of the difference in soft color ($S_\mathrm{HR1}$) reduces from 8.75 to 3.47. The significance parameter of hard color ($S_\mathrm{HR2}$) also reduces from 8.37 to 4.76. M82 X-1 still seems to undergo spectral variation during the other ACIS observations (Obs. 1, 2, 6 and 7). The color-luminosity diagram and the spectra reveal a transition from a low/hard state to a high/soft state. Detailed analysis and discussion of the spectral properties of M82 X-1 were presented by Agrawal and Misra (2006). Another spectral variable, CXOU J095550.7+694044, showed a inverse color-luminosity correlation to that of M82 X-1. In fact, CXOU J095550.7+694044 is a luminous young X-ray supernova remnant candidate near the center of M82, we will discuss it more in the next subsection. The other spectral variable we found in M82, CXOU J095546.6+694041, did not show a correlation between luminosity and color clearly.

\subsubsection{Supernova remnants}

CXOU J095546.7+694038, a source which is at the position of a historical supernova SN 2004am (Singer et al.~2004; Mattila et al.~2004; Beswick et al.~2004). SN 2004am was discovered in M82 on March 5, 2004 by Lick Observatory Supernova Search (LOSS) and the explosion is believed to take place at the end of 2003. SN 2004am was determined as a type II supernova. Although the positions of the X-ray source CXOU J095546.7+694038 and SN 2004am match well, this X-ray source has been detected before the supernova explosion. It suggests that CXOU J095546.7+694038 is not the X-ray counterpart of SN 2004am. However, the lightcurve of CXOU J095546.7+694038 (Figure 3) shows that in the observations after the supernova explosion, its X-ray luminosity increased by a factor of $\sim$ 50\%. It is not clear whether the increase of X-ray luminosity is due to the appearance of the supernova remnant of SN 2004am along the line of sight to the original X-ray source or not. We also check the \chandra\ spectra before and after the supernova event, the quality of these spectra is not good enough for us to explore the nature of the source.

CXOU J095550.7+694044, an ultraluminous SNR candidate near the center of M82, has a strong radio counterpart known as 41.95+575 (McDonald et al.~2001, 2002; Fenech et al.~2008). It has been suggested to be a young supernova event taking place within a high density molecular cloud (McDonald et al.~2001). The X-ray spectral property of CXOU J095550.7+694044 has also been discussed in the work of Kong et al.~(2007). In the archival \chandra\ ACIS data sets we used (observation 1, 2, 5, 6 and 7), it showed a spectral variation in $\textrm{HR2}$. The X-ray color of CXOU J095550.7+694044 seems softer in the observations when CXOU J095550.7+694044 is relatively dim.

CXOU J095552.8+694045 is also located within the error circle of a radio-selected supernova remnant in the catalog of the deep MERLIN 5GHz radio observation (Fenech et al.~2008). The flux of this X-ray source is close to the detection limit and the spectra suffer from strong contamination due to the diffuse emission. It is not clear whether it is the X-ray counterpart of the supernova remnant or not.

\section[]{Discussion}

Long-term flux variability of individual sources could result from various kinds of physics and structure of the sources, revealing information about their physical nature. However, here we are concerned more about the source variability of the whole galaxy. Our study shows that there is an prominent fraction of transients and variables in M82. We are therefore interested in some questions: What kind of source population and mechanism could explain the long-term variability of M82? Is the source variability of star-forming galaxies higher than that of other types of galaxies? We believe that the overall variability of a galaxy reflects the nature of its underlying source population. Considering that previous studies of X-ray sources in external galaxies are hitherto mainly based on X-ray luminosity functions (XLFs), can we further constrain the source populations with the information of temporal properties? In this section, we will first discuss the source population and the possible origins of the long-term variability of M82. And then we will further compare the long-term variability of different morphological types of galaxies.

Luminous X-ray sources in galaxies are believed to be dominated by XRBs. Based on previous studies of XLFs and other characteristics of our Galaxy and different types of external galaxies (see a review article by Fabbiano 2006), LMXBs are thought to dominate bright X-ray sources in early type galaxies. HMXBs dominate the source populations of galaxies with high star formation rate (SFR) to total stellar mass ratio. For spiral galaxies, the populations are more complex with a combination of both LMXBs and HMXBs, which differs from bulge, disk to arms. Providing the strong star-forming activities in M82, HMXBs associated with young stellar populations are expected to dominate the bright X-ray sources we detected. With XLFs of galaxies, it is possible to discriminate the contribution of HMXBs and LMXBs. The XLFs of HMXBs show a flatter slope of power law distribution without high luminosity cut-off. In contrast, high luminosity cut-off is seen in XLFs of LMXBs (Grimm et al. 2002). Grimm et al. (2003) further studied XLFs of star-forming galaxies. After excluding the estimated contribution from LMXBs, they suggested that there is a universal XLF of HMXBs characterized by a power law distribution with a cumulative slope of $-0.61\pm0.12$. Here we present the XLFs of M82 in all the observations we used (Figure 4). Except for observation 6 and 7, the XLFs cover the sources within nearly all the $D_{25}$ isophote of M82. The low luminosity part of the XLFs of M82 is highly incomplete due to its strong diffuse emission. We set a completeness threshold at $10^{38}$\lum and then fit the differential XLFs with power law using maximum likelihood. The cumulative slopes of all observations (except those with incomplete coverage of the $D_{25}$ isophote) of M82 are shown in Table 3. The XLF of M82 shows a flat power law distribution without high luminosity cut-off, suggesting a dominating HMXB population in the system. Furthermore, the good match of the cumulative slope and that of the universal XLF of HMXBs suggests that the contribution from LMXBs is fairly low. Therefore, we consider most of the sources in M82 to be HMXBs, which can constrain the origins of the long-term variability we detected. Apart from XLFs, we also consider a classification scheme developed by Prestwich et al. (2003) using X-ray colors of sources. However, the high absorption of M82 makes it more difficult to probe the source population using X-ray colors.

\begin{figure}
\centering
\includegraphics[width=6.2cm,angle=270]{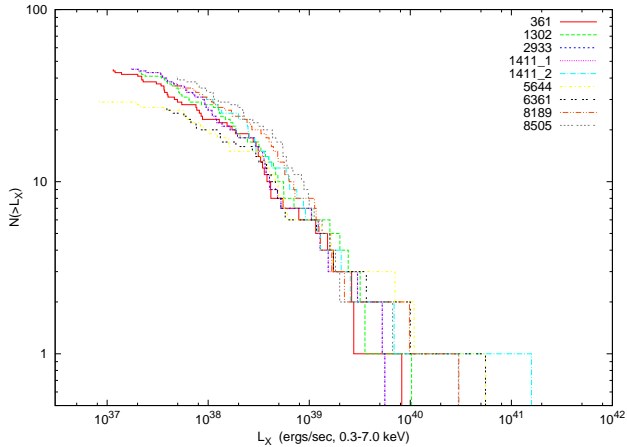}
\caption{Cumulative X-ray luminosity functions of M82 in all observations. The observations cover nearly all the $D_{25}$ isophote of M82 except observation 6 and 7.}
\end{figure}

\begin{table}
\centering{\scriptsize
\caption{Cumulative slopes of the XLFs of M82}
\begin{tabular}{cccccc}
\hline
\hline
No. & Obs.~ID & Slope \\
\hline
1	&	361	& $-0.72$\\
 
2	&	1302	& $-0.65$\\

3	&	1411-1	& $-0.64$\\

4	&	1411-2	& $-0.65$\\

5	&	2933	& $-0.65$\\

8	&	8189	& $-0.63$\\

9	&	8505	& $-0.63$\\

\hline
\end{tabular}
}
\par
\medskip
\begin{minipage}{0.95\linewidth}
NOTE.--- Observation 6 and 7 are not included because of the incomplete coverage of the $D_{25}$ isophote of M82 .
\end{minipage}
\end{table}

In addition, luminosity level reached by sources could impose constraint on source population since a higher observed luminosity of a source may imply a higher mass of its compact object if the Eddington limit is not violated. Therefore, luminosities above $10^{39}$\lum reduce the possibility for an XRB to host a neutron star. In fact, about half of the X-ray sources we detected in M82 exceed the Eddington limit of typical neutron stars (a few times $10^{38}$\lum), suggesting that there may be half or more BH XRB systems in M82. This statement can be further supported by the studies of source colors. Based on X-ray colors, Colbert et al. (2004) suggested that the dominant sources in nearby star-forming galaxies are generally inconsistent with accretion-powered NS HMXBs and more consistent with softer BH HMXBs. Besides, Swartz et al. (2004) showed that the properties of ULXs are similar to those of fainter sources, implying that ULXs are the high luminosity end of HMXBs, supernovae and LMXBs. It is therefore improper to treat ULXs as a distinct class of X-ray sources. In summary, we believe that the bright X-ray sources in M82 are dominated by HMXBs, and more than half of them should be black hole systems.

Based on the above discussion about source population, we could further constrain the possible origins of the detected source variability into temporal phenomena related to HMXBs. Among different types of XRBs, various temporal properties have been observed and to some extent explained by theories. For Roche lobe overflow systems, BH LMXBs in the Galaxy are more variable than neutron star systems (see the X-ray variability functions of Galactic source population in Figure 5 of Feng and Kaaret 2006). XRBs with a neutron star or a high mass stellar companion are thought to be relatively stable. Irradiation of the accretion disk resulted from the emission of a young massive donor star or the surface of a neutron star can maintain a temperature higher than the hydrogen ionization temperature, $T_{\rm{H}} \sim 6500$ K, thereby suppressing the instabilities in its accretion disk (e.g. van Paradijs 1996; Irwin 2006). On the other hand, HMXBs could also be transients. In our Galaxy, more than half of the HMXBs are systems consist of a Be star and a neutron star. The rapid rotation of the Be star created an extended circumstellar disk. If the orbit is sufficiently eccentric, accretion could be triggered all of a sudden when the compact object pass through the disk. Therefore, almost all the Be HMXBs are transients. For HMXB dominant galaxies like M82, many transients should be Be HMXBs. Note that the compact objects of Be HMXBs in our Galaxy are mainly neutron star, but more black hole systems are expected in M82. This does not oppose our inference. Considering the formation of Be HMXBs, the chance to form neutron star systems is fairly higher than that of black hole systems. Therefore, the presence of some luminous black hole systems may imply a large amount of neutron star systems with lower luminosities.

Based on strong source variability, we infer that many of the sources in star-forming galaxies are Be transients. This inference could be checked from other aspects. In a HMXB, mass is accreted through either stellar wind or Roche lobe overflow. For both the stellar wind systems and the Roche lobe overflow systems, the mass donor could be an OB-supergiant or a Be star. From a theoretical study (Postnov 2003) of the universal XLF of HMXBs deduced by Grimm et al. (2003), the $-0.6$ cumulative slope of XLF could be explained by the mass-luminosity and mass-radius relations of wind-fed HMXBs. On the other hand, the slope of Roche lobe overflow systems is not consistent with the observed universal slope. Postnov (2003) considered that for HMXBs with Roche lobe overflow, the high efficiency supercritical accretion produces luminosities on the order of the Eddington limit, which makes their XLF close to the mass distribution of the compact objects. The different XLF slope with respect to the $-0.6$ universal value suggests that Roche lobe overflow population may not make a significant contribution. After limiting the major contribution to the universal XLF in wind-fed HMXBs, we could further reject the wind-fed systems with a donor of supergiant, because they possess relatively lower luminosities ($10^{35}$ $-$ $10^{36}$\lum) than the detected sources in our study. For these reasons, we exclude the significant contribution of Roche lobe overflow systems and the fainter wind-fed supergiant systems. The major population of the X-ray sources in M82 and the observed universal HMXB XLF is most likely to be wind-fed Be HMXB. This is consistent with our previous inference based on source variability.

Next, we discuss long-term source variability of galaxies with different morphological types. Not many studies about this have been done up to now. With our results of the prototype star-forming galaxy M82 and previous studies of several galaxies, a rough picture could be constructed.

The first way to compare source variability of different systems is based on X-ray variability function (XVF). Feng and Kaaret (2006) studied the temporal properties of bright X-ray sources in two extremely different host environments: the old elliptical galaxy NGC 1399 and the young, star-forming Antennae galaxies (NGC 4038/4039). Maximum ratios in luminosity, $R$, of sources were obtained using similar analysis process of our work. However, it should be noted that the luminosity ratios here are all lower limits. For both of their two target galaxies, they used 7 \chandra\
 ACIS observations within 5 years. Here we present our X-ray variability function of M82 combined with the results of Feng and Kaaret's work (Figure 5). A luminosity cut-off of $3\times10^{38}$\lum\ is set since the detection limits in most of the observations of these three galaxies are below this level. The early type galaxy, NGC 1399, hosts many bright X-ray sources, and their variability is clearly lower than the other two starburst systems. For early type galaxies, bright X-ray sources are dominated by LMXBs. Although many LMXBs in our Galaxy show transient phenomena, the bright sources in NGC 1399 are relatively stable with respect to those young HMXBs in star-forming galaxies. However, we expect that the variability of LMXBs could be more prominent on longer timescale.

Compared with the Antennae galaxies, M82 hosts more X-ray sources beyond the threshold and more high $R$ sources. However, it is premature to claim that the source variability of M82 is intrinsically higher than that of the Antennae. Because of the lower detection limit of M82, we can detect variations with higher $R$. It is possible that the detailed source populations and source variability are slightly different in star-forming galaxies with different conditions. To have a better understanding of the relation between X-ray variability in star-forming environment and star-formation rate (SFR), star-formation history and/or other factors, more studies are required.

\begin{figure}
\centering
\includegraphics[width=8.6cm,angle=0]{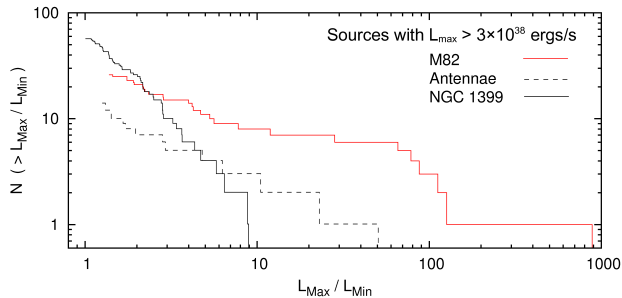}
\caption{X-ray variability function of bright sources in M82 combined with Feng and Kaaret's results (Figure 5 in Feng and Kaaret 2006) of the old elliptical galaxy NGC 1399 and the young, star-forming Antennae galaxies (NGC 4038/4039).}
\end{figure}

We suggest another way to compare source variability of galaxies. As we mention in \S\,4.2, the fraction of variables with $R > 3$ could be an indicator of variability. X-ray variability functions plotted with the lower limit of $R$ give a detailed picture of variability distribution. However, with current instruments, detection limits of external galaxies outside the local group are not enough to cover the sources in their low states. Therefore, the XVFs, especially the high $R$ part of them strongly depend on the detection limits. XVFs should only be used to compare systems with similar detection limits. To compare the few studies done with different detection limits, the fractions of variables with $R > 3$ provide a rough but proper description.

Two systems with moderate star-forming activities, NGC 6946 and NGC 4485/4490, were well studied with an emphasis on long-term variability by Fridriksson et al. (2008). The detailed data analysis and the detection limits (a few times $10^{36}$\lum\ for NGC 6946 and $\sim 10^{37}$\lum\ for NGC 4485/4490) are not exactly the same as ours but comparable. We did not compare their results with ours using XVF since there are relatively few bright sources beyond the threshold we set in the previous comparison. We recalculated the fractions of variables with $R > 3$ of these two systems using the criteria we used in this paper. We obtained 23\% for NGC 6946 and 21\% for NGC 4485/4490. The long-term variability of the X-ray sources in M82 (47\% sources are variables with flux ratio $> 3$) is extremely high compared to these two galaxies with star-forming activities. Since the SFRs in these two galaxies are lower than that of M82 (see the SFRs of galaxies in Table 4), it is reasonable to claim that SFR may be a correlated factor of long-term X-ray variability. This is also consistent with the concern that there is a causality between source populations and temporal properties since SFR is an important indicator of source populations.

\begin{table}
\centering{\scriptsize
\caption{Inferred properties of host galaxies}
\begin{tabular}{cccccc}
\hline
\hline
Galaxy & SFR$^{a}$ ($M_\odot$ $yr^{-1}$ ) & Mass$^{b}$ ($M_\odot$)\\
\hline
M82	&	9.3	&	$6.3\times10^{10}$\\

Antennae	&	8.3	& $5\times10^{10}$\\

NGC 1399	&	...	&	$1.6\times10^{11}$\\

NGC 6946	&	4	&	$2\times10^{11}$\\

NGC 4485$/$4490	&	4.7	&	$1.8\times10^{10}$\\

\hline
\end{tabular}
}
\par
\medskip
\begin{minipage}{0.95\linewidth}
$^a$ M82, Antennae (Colbert et al. 2004), NGC 6946 (Sauty et al. 1998) and NGC 4490 (Clemens et al. 1999).

$^b$ Stellar mass of M82, Antennae and NGC 1399 were calculated using $L_{\rm{k}}$ (Colbert et al. 2004). Dynamical mass is    	used in NGC 6946 (Crosthwaite and Turner 2007) and NGC 4485/4490 (Clemens et al. 1999).

\end{minipage}
\end{table}

\section[]{Conclusions}

We have reported a study of the long-term variability of the X-ray sources in the starburst galaxy M82. By combining 9 archival \chandra\ data sets observed between 1999 and 2007, we detected 58 X-ray point sources within the $D_{25}$ isophote of M82 down to a luminosity of $\sim 10^{37}$\lum. Seven detections can be expected from background sources. Three sources are coincident with the positions of historical supernovae or radio-selected SNRs.

Twenty-six sources in M82 exhibit long-term (i.e., days to years) flux variability and 3 exhibit long-term spectral variability. Furthermore, we classified 26 sources as variables and 10 as persistent sources. Among the total 26 variables, 17 varied by a flux ratio of $> 3$ and 6 are transient candidates. To reduce the selection effects, we used the fraction of variables with a flux ratio of $> 3$ to be an indicator of long-term variability. The long-term variability of the X-ray sources in M82 is higher than Early-type galaxies and some galaxies with star-forming activities (NGC 6946 and NGC 4485/4490).

For the variables, transient candidates and spectral variables we classified, various kinds of variability properties are shown. Two of the transients showed recurrent outbursts. One of the variables was classified as SSS. M82 X-1 showed a transition from a low/hard state to a high/soft state in its color-luminosity diagram, and another spectral variable showed an inverse correlation.

Based on the strong variability, the X-ray luminosity functions, and the number of ULXs, we suggest that the X-ray source population of M82 with luminosities greater than $\sim 10^{37}$\lum\ is dominated by HMXBs. More black hole XRBs should be expected than neutron star XRBs. Some of the transients should be Be HMXBs.

\section*{Acknowledgments}

This project is supported by the National Science Council of the
Republic of China (Taiwan) through grant NSC96-2112-M-007-037-MY3.

\newcommand{\iaucirc}{{IAU Circ.}}
\newcommand{\aj}{{AJ}}
\newcommand{\mnras}{{MNRAS}}
\newcommand{\apj}{{ApJ}}
\newcommand{\araa}{{ARA\&A}}
\newcommand{\apjs}{{ApJS}}
\newcommand{\aap}{{A\&A}}
\newcommand{\apjl}{{ApJL}}

%\bsp

\label{lastpage}

\end{document}